\documentclass[10pt, conference]{ieeeconf}      

\usepackage{graphicx}%
\usepackage{multirow}%
\usepackage{amsmath,amssymb,amsfonts}%
\usepackage{mathrsfs}%
\usepackage{xcolor}%
\usepackage{textcomp}%
\usepackage{manyfoot}%
\usepackage{booktabs}%
\usepackage{algorithm}%
\usepackage{algorithmicx}%
\usepackage{algpseudocode}%
\usepackage{listings}%
\usepackage{cite}

\usepackage{pifont}
\usepackage{subfig}    
\usepackage{marvosym}
\usepackage{url} 
\usepackage{tikz}
\newcommand*\circled[1]{\tikz[baseline=(char.base)]{
            \node[shape=circle,draw,inner sep=1pt] (char) {#1};}}

\IEEEoverridecommandlockouts 
\title{\LARGE \bf
Towards Robust Blockchain Price Oracle: A Study on Human-Centric Node Selection Strategy and Incentive Mechanism
}

\author{Youquan Xian, Xueying Zeng, Hao Wu, Danping Yang, Peng Wang and Peng Liu\textsuperscript{\Letter}
\thanks{*This work was supported in part by the Guangxi Science and Technology Major Project (No. AA22068070), the National Natural Science Foundation of China (Nos. 62166004,U21A20474), the Basic Ability Enhancement Program for Young and Middle-aged Teachers of Guangxi (No. 2023KY0062), Innovation Project of Guangxi Graduate Education (Nos. XYCBZ2024025)}
\thanks{Youquan Xian, Xueying Zeng, Hao Wu, Danping Yang, Peng Wang and Peng Liu are affiliated with both the Key Lab of Education Blockchain and Intelligent Technology, Ministry of Education, and the Guangxi Key Lab of Multi-Source Information Mining and Security at Guangxi Normal University, Guilin, 541004, China. \textsuperscript{\Letter} Corresponding author: Peng Liu (\tt\small liupeng@gxnu.edu.cn)}%
}

\begin{document}
\maketitle
\thispagestyle{empty}
\pagestyle{empty}

\begin{abstract}
As a trusted middleware connecting the blockchain and the real world, the blockchain oracle can obtain trusted real-time price information for financial applications such as payment and settlement, and asset valuation on the blockchain. However, the current oracle schemes face the dilemma of security and service quality in the process of node selection, and the implicit interest relationship in financial applications leads to a significant conflict of interest between the task publisher and the executor, which reduces the participation enthusiasm of both parties and system security. Therefore, this paper proposes an anonymous node selection scheme that anonymously selects nodes with high reputations to participate in tasks to ensure the security and service quality of nodes. Then, this paper also details the interest requirements and behavioral motives of all parties in the payment settlement and asset valuation scenarios. Under the hypothesis of rational man, an incentive mechanism based on the Stackelberg game is proposed. It can achieve equilibrium under the pursuit of the revenue of task publishers and executors, thereby ensuring the revenue of all types of users and improving the enthusiasm for participation. Finally, we verify the security of the proposed scheme through security analysis. The experimental results show that the proposed scheme can reduce the variance of obtaining price data by about 55\% while ensuring security, and meeting the revenue of all parties.

\end{abstract}


\section{Introduction}


As one of the most prominent applications in blockchain, Decentralized Finance (DeFi) is thriving, supported by its characteristics of decentralization, openness, and inclusiveness, bringing profound changes and innovations to the financial sector \cite{kitzler2023disentangling}. The emergence of price oracles fills the information gap in decentralized financial systems, injecting trustworthy external world data into the DeFi ecosystem, such as asset prices and interest rates, enabling DeFi applications to utilize these data for various financial activities, such as payment settlements and asset evaluations \cite{zhao2022toward}.

With the development of blockchain, oracle technology has attracted widespread research attention, especially in the design of node selection strategies and incentive mechanisms. Among these, node selection schemes in oracles can generally be categorized into two types: those based on Verifiable Random Function (VRF)\cite{micali1999verifiable} and those based on reputation. For instance, schemes such as DOS Network \cite{dos} and Lin et al. \cite{lin2022novel} used the unpredictability and verifiability of VRF to ensure the anonymity security of the node selection process, thereby avoiding nodes being targeted by adversaries and returning incorrect data. In contrast to the pursuit of security in the aforementioned schemes, reputation-based node selection strategies prioritize the service quality of selected nodes to enhance user experience. Goswami et al. \cite{goswami2022towards} argued that randomly selecting oracle nodes to perform a task may be challenging to obtain an optimal solution. Therefore, the authors propose a middleware that selects the best data provider available in the Chainlink oracle network \cite{chainlink} based on the service requirements of individual tasks. Similarly, Xian et al. \cite{xian2024distributed} selected nodes that respond quickly to tasks based on potential matching relationships between data sources and oracle nodes to improve system response time. Additionally, the authors introduced a new oracle architecture to provide trustworthy device information for node selection strategies. Taghavi et al. \cite{taghavi2023reinforcement} evaluated nodes through lightweight reinforcement learning methods, efficiently selecting the most reliable and economical nodes to execute tasks, thereby improving system service quality and user experience.

Blockchain oracle, as a distributed system, relies on a large and actively engaged user base to ensure the system's resilience against attacks and long-term stability \cite{pasdar2023connect}. However, existing research on oracle incentive mechanisms primarily focuses on incentivizing oracle nodes, namely task executors. Early oracle solutions such as Augur \cite{peterson2015augur} and Astraea \cite{berryhill2019astraea} primarily incentivize honest behavior among task executors to ensure their active participation and the authenticity of submitted data. Subsequently, Cai et al. \cite{cai2020truth, cai2022truthful} and Nelaturu et al. \cite{nelaturu2020public} mitigated herd behavior \cite{galariotis2015herding} by associating rewards with voting shares, thus discouraging executors from blindly selecting the same data as the majority for profit. Chainlink, as a prominent commercial oracle solution, primarily employs a reputation-based mechanism that evaluates nodes' historical service levels to allocate higher rewards to high-reputation nodes, incentivizing them to maintain availability and data integrity \cite{chainlink}.

Although Du et al. \cite{du2022novel} identified that current incentive mechanisms fail to consider the revenue of both task publishers and executors simultaneously, leading to potential disadvantages for one party when unilaterally incentivizing the other. They proposed an incentive mechanism based on auction theory to address the revenue of both parties and thereby enhance multi-party participation. However, in practical applications, it is difficult to accurately estimate the cost of obtaining data for different users. Furthermore, the implicit interest relationship in financial applications will drive the interest-related oracle nodes to upload wrong data to seek improper benefits. Conflicting revenue between task publishers and executors in price oracles reduces the enthusiasm of both parties for participation. 
Therefore, to enhance the robustness of price oracles, it is necessary to: \ding{182} address the dilemma between anonymity security and service quality in the node selection process, and \ding{183} design incentive mechanisms that consider the revenue of multiple parties for DeFi applications with significant conflicts of revenue such as payment settlement and asset valuation. Despite extensive research on blockchain oracles, there is no comprehensive solution to the aforementioned issues.


In this paper, we first integrate reputation mechanisms with cryptographic designs to design a node selection strategy. It ensures the anonymity of nodes and enhances user experience by randomly selecting high-quality oracle nodes. Furthermore, we conduct a detailed analysis of the behavioral motivations of both task publishers and executors in price oracle systems used for payment settlement and asset valuation. Under the hypothesis of rational man, we design a new incentive mechanism based on the Stackelberg games. The objective of this mechanism is to achieve an equilibrium solution that satisfies the revenue of both task publishers and executors simultaneously, enhancing user participation.

The contributions of this paper are summarised as follows:
\begin{itemize}
\item To balance both the security and service quality of oracles, we integrate a reputation system with a cryptographic design to design a secure node selection strategy. It ensures both node anonymity and enhances the quality of data acquisition.
\item We analyze the revenue and behavioral motivations of various participants in price oracle systems and construct them into a Stackelberg game model under the hypothesis of rational man. The equilibrium solution obtained simultaneously satisfies the revenue of both task publishers and executors, to enhance user participation.
 \item The security analysis verifies the security of the proposed scheme. The experimental results show that our scheme can improve the quality of the acquired data while ensuring the anonymity of the nodes, and the equilibrium solution of the proposed incentive mechanism is also the optimal solution for the participants.
\end{itemize}

The rest of this paper is structured as follows: Section \ref{work} details the proposed scheme. Section \ref{result} analyzes the effectiveness of the proposed scheme through experiments. Section \ref{security_analysis} analyzes the security of the proposed scheme. Section \ref{conclusion} concludes the paper and outlines some future work ideas.

\section{System Model}
\label{work}

\begin{figure}[h]
  \centering
  \includegraphics[width=\linewidth]{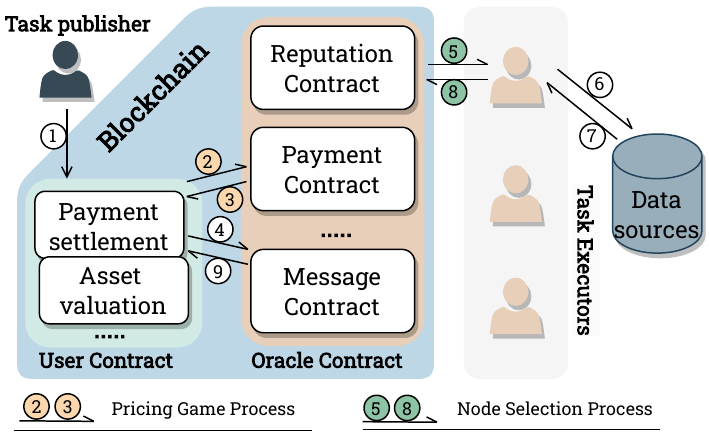}
  \caption{Overview of the proposed oracle system.}
  \label{fig:overview}
\end{figure}

\subsection{System Flow}
As shown in Fig. \ref{fig:overview}, the proposed price oracle solution focuses on the node selection process and the crucial game-theoretic pricing design in the incentive mechanism. The system workflow is outlined as follows:

\circled{1} Users (task publishers) invoke the user contract which embedded oracle interface, such as the transaction payment contract; \circled{2}-\circled{3} The user contract calls the oracle contract interface for fee estimation, obtaining the most reasonable service fee valuation $P$ based on the transaction commodity scale $K$ and $\alpha$. Here, $\alpha$ is a hyperparameter representing the degree to which the publisher values data quality. A higher $\alpha$ indicates that the publisher values data quality more than the expenses incurred. We will elaborate on it in Section \ref{stackelberg}. \circled{4} After pledging the service fee $P$ to the oracle contract, the user contract initiates a data request to the oracle contract, and the oracle contract records the request event $E=(Q,D,P,K)$ on the blockchain. Here, $Q$ serves as the unique identifier for $E$, and $D$ represents the set of data sources needed.
\circled{5} Upon observing a new request event $E$ on the blockchain, oracle node (task executor) $O_i$ computes its own random number $R_i$ and proof $\chi_i$ using its private key $sk_i$ and the current blockchain's random number $R$. $O_i$ then determines whether it has been selected to participate in this task based on its computed $R_i$ and the optional range $\Gamma_i$ queried from the reputation contract. The details of node selection will be discussed in Section \ref{node_select}. \circled{6}-\circled{7} If node $O_i$ is secretly selected, it can randomly choose a data source from the set $D$ and request to obtain data $X_i$.
\circled{8} Node $O_i$ submits data based on the commit-reveal protocol. Firstly, in the commit phase, it uploads the credential $\Pi = (\chi_i,R_i, \xi_i)$. ($\chi_i$,$R_i$) are used to verify whether the node is selected, and $\xi_i = Hash(X_i \parallel pk_i)$ serves as the credential for data $X_i$. Then, after the commit phase ends, it uploads the data $X_i$ and $pk_i$ to prevent the issue of Freeloading\cite{chainlink}.
\circled{9} The message contract first filters the data, with values having greater bias more likely to be removed, and then aggregates the remaining data. Subsequently, the reputation contract updates the node's reputation value $C_i$ and the optional range $\Gamma_i$ based on the quality of the data uploaded by node $O_i$. The payment contract then evenly distributes the reward $P$ among the nodes that successfully aggregate the data. Finally, the message contract feeds back the final price data to the user contract through a callback mechanism for the user contract to proceed with subsequent transaction processes.

\subsection{Anonymous Node Selection}
\label{node_select}

The reputation contract evaluates the reputation value of a node based on its historical service level, as seen in schemes such as DAON \cite{dong2023daon} and Razor \cite{huilgolkar2021razor}. In this paper, we measure the service quality of nodes based on the consistency of the data they upload. In the proposed approach, a high reputation increases the probability of being selected. The reputation value $C_i$ of any oracle node $O_i$ is expressed as follows:

\begin{eqnarray}
    C_i = e^{- d(X_i, \mu)}
\end{eqnarray}

$X_i, \mu$ represents the price returned by the node $O_i$ and the average of all upload prices. $d(x,y)$ is a function that describes the distance between $x,y$, such as the Manhattan distance used in this article.

Upon receiving the request event $E$, off-chain oracle nodes query the reputation contract for their optional range $\Gamma_i$ to participate in this task. $\Gamma_i$ is generated by the reputation contract based on the reputation value $C_i$ of the node and the expected number of nodes to be selected for this task, denoted as $M$. Its significance lies in adjusting the number of nodes selected per round based on the expected $M$.

\begin{eqnarray}
   \Gamma_i = C_i \times \frac{M}{\sum C_i}
\end{eqnarray}

Then, the oracle node $O_i$ uses VRF \cite{micali1999verifiable} to generate a random number $R_{i}$ and it's proof $\chi_i$ based on its private key $sk_i$ and the global random number $R$. Where $R$ can be a hash of the current latest block or a verifiable random number, $ 2^{256} $ is the maximum value of the random number generated by SHA256 in VRF.

\begin{eqnarray}
   R_i = \frac{VRF_{generate}(R,sk_i)}{2^{256}}
\end{eqnarray}

\begin{figure}
\centering
  \includegraphics[width=\linewidth]{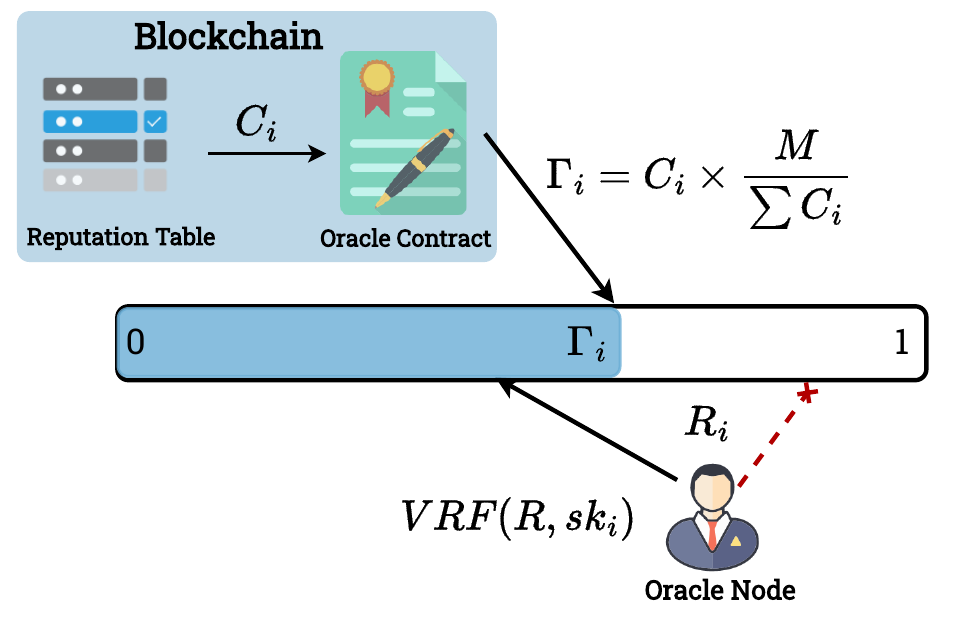}
  \caption{Anonymous node selection.}
  \label{fig:select}
\end{figure}

Finally, as shown in Fig. \ref{fig:select}, the node $O_i$ determines whether it is selected based on the locally calculated $R_i \in [0,1)$ and the optional range $\Gamma_i$ of $O_i$ in the reputation contract, where $S_i = 1$ represents $O_i$ being selected for this task.

\begin{eqnarray}
S_i = 
\begin{cases}
 1, & \text{if} \ R_i \leq \Gamma_i\\
 0, & \text{if} \ R_i > \Gamma_i
\end{cases}
\end{eqnarray}

When verifying, it is only necessary to verify whether the node is selected to participate in the task according to $\chi_i$, $R_i$, $R$, $pk_i$.

\begin{eqnarray}
   0/1 \gets VRF_{verify}(R_i,\chi_i,R,pk_i)
\end{eqnarray}

\subsection{Incentive Mechanism}
\label{stackelberg}

Taking payment settlement and asset valuation as examples, both the buyer and the seller, as task publishers, need to obtain trustworthy real-time prices to conduct transactions. Their goal is to spend less on oracle service fees $P$ while obtaining trustworthy off-chain asset prices $X$. However, due to real-time price fluctuations, the price $X_i$ obtained by task executor $O_i$ may differ. If there is some form of vested interest between task executor $O_i$ and one party in the transaction, this executor can upload a modified price $X_i^{'}$ to gain more profit. We refer to such executors as malicious executors.

\begin{eqnarray}
   V = K \times |X - X^{'}|
\end{eqnarray}

$V, K, X, X^{'}$ represents the improper profit of the executor, the number of goods traded, the price of the unmodified goods, and the price of the modified goods. It can be found that improper profit $V$ is related to the value of the executor's modified price $\Delta = |X_i - X_i^{'}|$.

In the average aggregation strategy, the final uploaded $n$ prices will be aggregated into a final price $X = \frac{\sum^{n}_{i=1} X_i}{n}$.
$G(\Delta)$ denotes the change in the final price $X$ by the $\Delta$ of a single executor, which can be expressed as $G(\Delta) = \frac{\Delta}{n}$.

To make malicious executors unable to modify their prices indefinitely to seek illegitimate revenue, we use a simple probability screening algorithm. Specifically, we let the data with greater deviation have a greater probability of being screened out. $h(\Delta) = e^{-\Delta}$ represents the probability that the data after modifying $\Delta$ will be adopted for aggregation, $h(\Delta) \in (0,1]$.

Under the condition of anonymous selection, malicious executors only know their situation, and the system selects nodes with relatively high reputations. Therefore, we assume that users follow the Minimax strategy \cite{aumann1972some}. Each malicious executor estimates their minimum earnings, assuming that all other selected nodes except themselves are honest nodes and obtain consistent data, i.e., sharing the service fee $P$ equally. Therefore, $\frac{P}{n}$ and $K \times G(\Delta)$ respectively represent the earnings of the executor when behaving honestly and maliciously. The total revenue function $U_2$ can be expressed as the sum of regular revenue $\frac{P}{n}$ and illegitimate revenue $K \times G(\Delta)$. However, the premise for the executor to receive revenue $U_2$ is successful aggregation rather than being filtered out.

\begin{eqnarray}
   U_2 = h(\Delta) \times (\frac{P}{n} +  \frac{K \times \Delta}{n} )
\end{eqnarray}

For publishers, their revenue function is related to the degree of deviation $\Delta$ in obtaining data and the service fee $P$ paid. When a node is excluded due to inconsistent data, its service fee will also be refunded to the task publisher. The hyperparameter $\alpha$ represents the publisher's expected target for data quality. Additionally, for convenience in subsequent calculations, we use the same $e^{-\Delta}$ as $h(\Delta)$ to evaluate the degree of data deviation.

\begin{eqnarray}
   U_1 = \alpha \times e^{-\Delta} + (1-\alpha) \times h(\Delta) \times (-P)
\end{eqnarray}

\begin{figure}[h]
  \centering
  \includegraphics[width=0.8\linewidth]{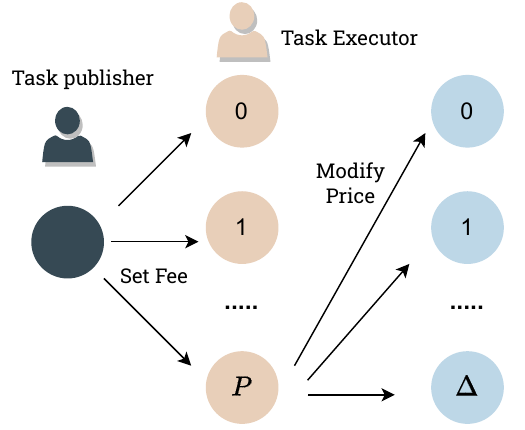}
  \caption{Game tree between publisher and executor.}
  \label{fig:game}
\end{figure}

To simultaneously satisfy the revenue of both parties and enhance user engagement, we model this problem as a Stackelberg game \cite{van2008overview}. We define the task publisher as the leader and the malicious executor as the follower. The leader makes the first move, and the follower makes decisions after observing the leader's actions. Both the leader and the follower have a set of available strategies, denoted as $S_1$ and $S_2$, respectively. $S_1 = \{P \}$ represents the service fee for the task, while $S_2 = \{\Delta\}$ represents the magnitude of the price modification chosen by the follower. For each set of strategies, the leader and follower respectively obtain a revenue value $U_1$ and $U_2$. Fig. \ref{fig:game} shows the dynamic decision game tree of the game.

To solve this problem using backward induction, we first maximize the follower's revenue $U_2$, i.e., finding the maximum value of $U_2$. We set $\frac{d U_2}{d \Delta} =0$ and $\frac{d^2 U_2}{d \Delta^2} < 0$, resulting in $\Delta = \frac{K-P}{K}$. Therefore, $\Delta^{*} = \frac{K-P}{K}$ represents the optimal decision for the follower given the service fee $P$ set by the leader. (See Appendix \ref{proof1} for details)

The leader substitutes $\Delta^{*}$ into their revenue function $U_1$ to find the optimal solution $P^{*}$. Following similar steps, we set $\frac{d U_1}{d P} = 0$ and $\frac{d^2 U_1}{d P^2} < 0$,  resulting in $P = \frac{K(\alpha-1) + \alpha}{(1-\alpha)}$, when $\frac{K}{1+K} \leq \alpha \leq \frac{2K}{1+2K}$. (See Appendix \ref{proof2} for details)

To conveniently adjust the hyperparameter $\alpha$, we design a scaling function $\Phi(\alpha)$ that maps $\alpha$ to the interval [0,1], while satisfying the constraints of $P$ and $\Delta$:

\begin{equation}
    \Phi (\alpha) = \frac{\alpha(1+K)(1+2K)}{K} - (1+2K)
\end{equation}

In summary, we have elaborated on the modeling and solving process of the game-based pricing strategy proposed in \circled{2}-\circled{3}. The oracle contract will recommend the service fee for this task based on the importance that the task publisher places on the task quality $\alpha$, and the quantity of traded goods $K$. The recommended service fee $P$ for this task is given by:

\begin{equation}
    P = \frac{K(\Phi (\alpha)-1) + \Phi (\alpha)}{(1-\Phi (\alpha))} 
\end{equation}

Under the hypothesis of rational man, both the task publisher and malicious task executors are expected to adopt the proposed recommendation to maximize their benefits.

\begin{table}[h] 
\centering
\caption{Experimental parameter settings.}
\label{tab:setting}
\begin{tabular}{|c|c|c|}
\hline
 Parameter & Meaning & Value \\ \hline
 $\lambda$ & The proportion of malicious nodes in the network & 0.4 \\ \hline
 $N$ & The number of nodes in the network & 50 \\ \hline
 $K$ & The number of goods to be traded in this task & 10 \\ \hline
 $M$ & The number of nodes to be selected in this task & 5 \\ \hline
 $\alpha$ & The importance of task publishers to data quality & 0.5 \\ \hline
\end{tabular}
\end{table}

\section{Evaluation}
\label{result}

\subsection{Experiment Setting}
We implement the prototype of the system. The on-chain part consists of several smart contracts written in Solidity. The off-chain part is implemented in Python and communicates with the on-chain contracts using web3.py. To ensure the authenticity of the experiments, we use the Truffle Suite \footnote{\url{https://trufflesuite.com/}} to generate an oracle network consisting of 50 active Ethereum blockchain nodes and deploy the smart contracts on a local Ethereum blockchain. To compare the service quality of oracles under similar security conditions, we set the Baseline to use VRF-based oracle schemes like DOS Network \cite{dos}. Finally, Table \ref{tab:setting} shows the parameters used in the experiments.

\begin{figure*}[h]
\centering
\subfloat[Data consistency]{\label{fig:var}\includegraphics[width=0.3\textwidth]{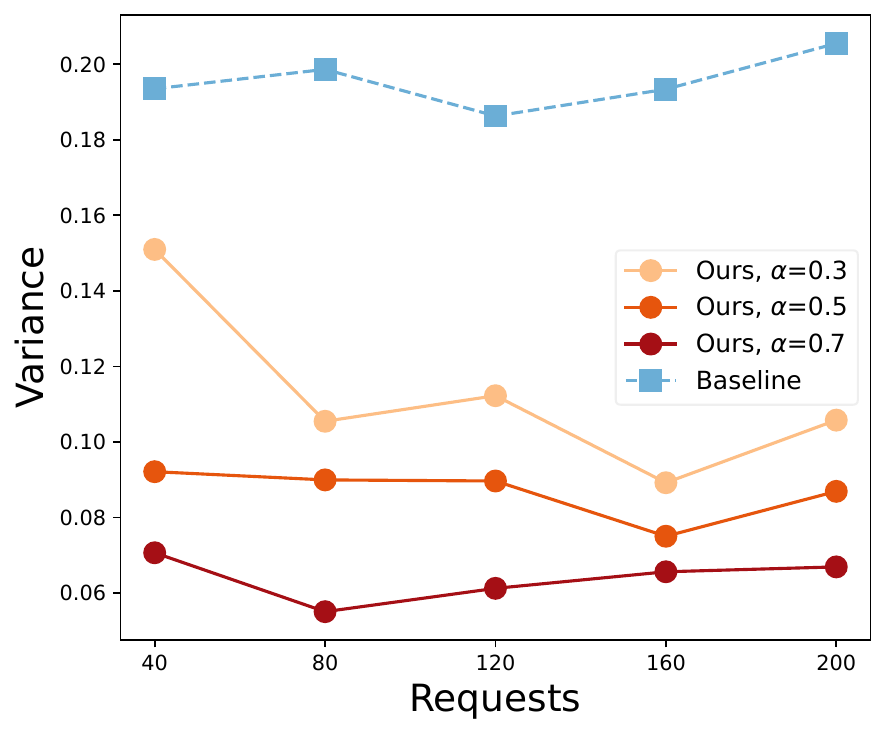}}\quad
\subfloat[Number of malicious node selections]{\label{fig:number_select}\includegraphics[width=0.3\textwidth]{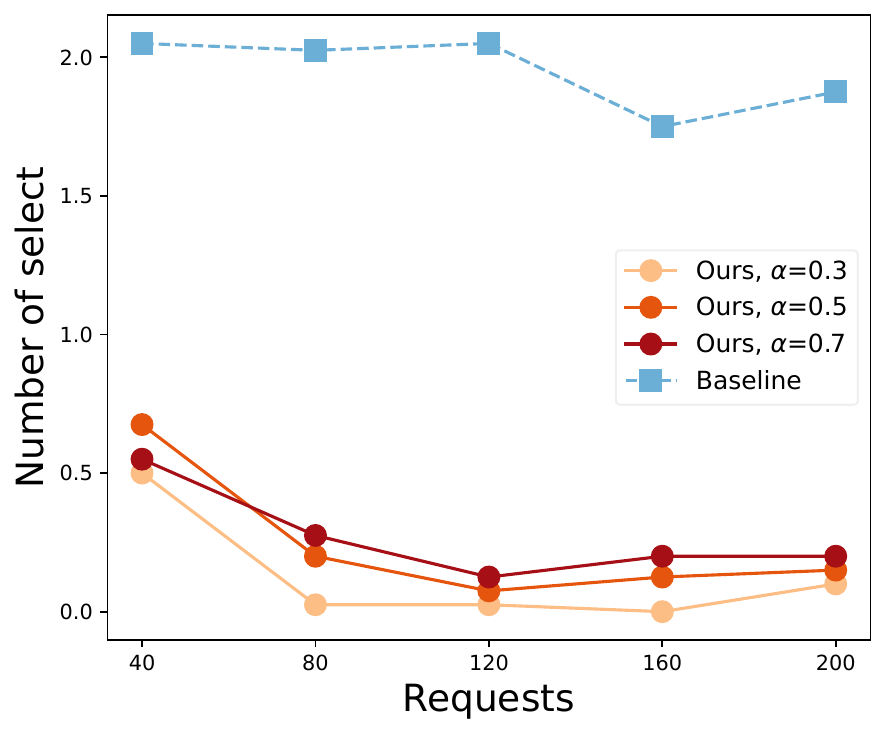}}\quad
\subfloat[Reputation of oracle nodes]{\label{fig:reputation}\includegraphics[width=0.3\textwidth]{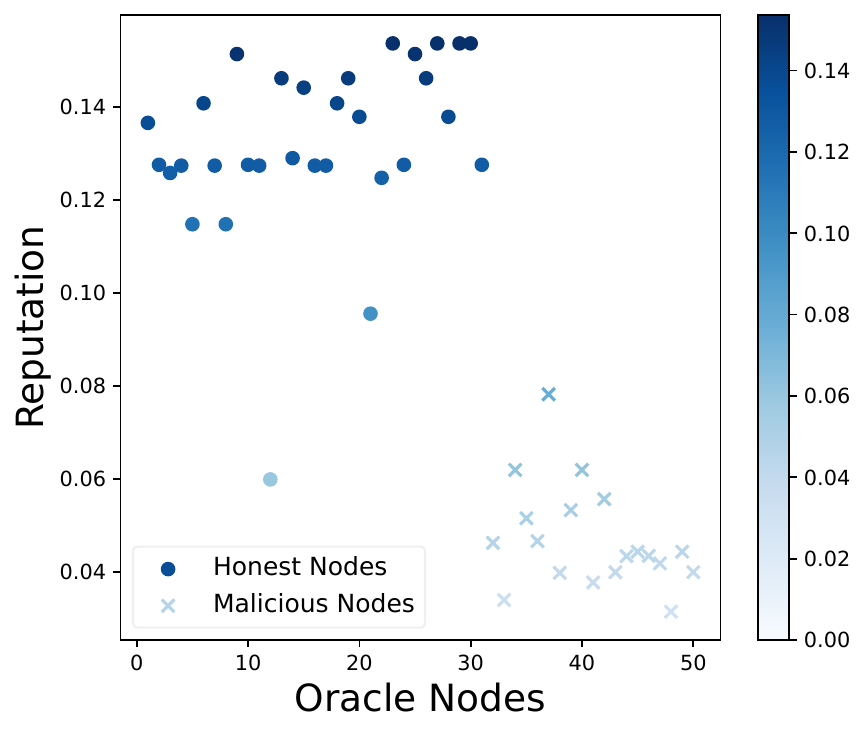}}\\
\caption{Compare node selection schemes.}
\label{fig:select_node}
\end{figure*}

\subsection{Quality of Service}

Fig. \ref{fig:var} compares the consistency between the proposed node selection scheme and the data obtained by Baseline. Here, we utilize data variance to denote data consistency. A smaller variance corresponds to greater consistency. It can be seen from the figure that our scheme can provide more consistent data for the blockchain, and with the increase of $\alpha$, the consistency of the obtained data will be higher. In particular, our scheme at $\alpha=0.5$ is about 55\% lower than the Baseline variance. It shows that the proposed node selection scheme can improve the quality of service of the system while continuing the anonymous security of VRF, and the higher the requirement of the task publisher for the quality of the acquired data, the higher the quality of the acquired data.

Fig. \ref{fig:number_select} shows the number of malicious nodes selected by different schemes for each task. We find that the proposed scheme can reduce the number of malicious nodes compared with the Baseline, which is the same as our original intention. The proposed scheme can improve the quality of system service by selecting high-quality nodes. However, we also find that with the increase of $\alpha$, the probability of selecting malicious nodes is also slowly increasing. This is because the incentive mechanism is in effect. With the increase of $\alpha$, the data uploaded by users is more consistent, which indirectly increases the difficulty of the reputation mechanism to distinguish malicious nodes.

Fig. \ref{fig:reputation} shows the reputation value of all nodes in the proposed scheme after 200 tasks. We can find that after multiple tasks, the reputation mechanism can identify the quality of service of nodes, and assign higher reputation values to honest nodes and lower reputation values to malicious nodes. This is also the reason why the proposed scheme can reduce the selection of malicious nodes.

\begin{figure}[h]
\centering
\subfloat[Benefits for task publishers]{\label{fig:u1}\includegraphics[width=0.45\linewidth]{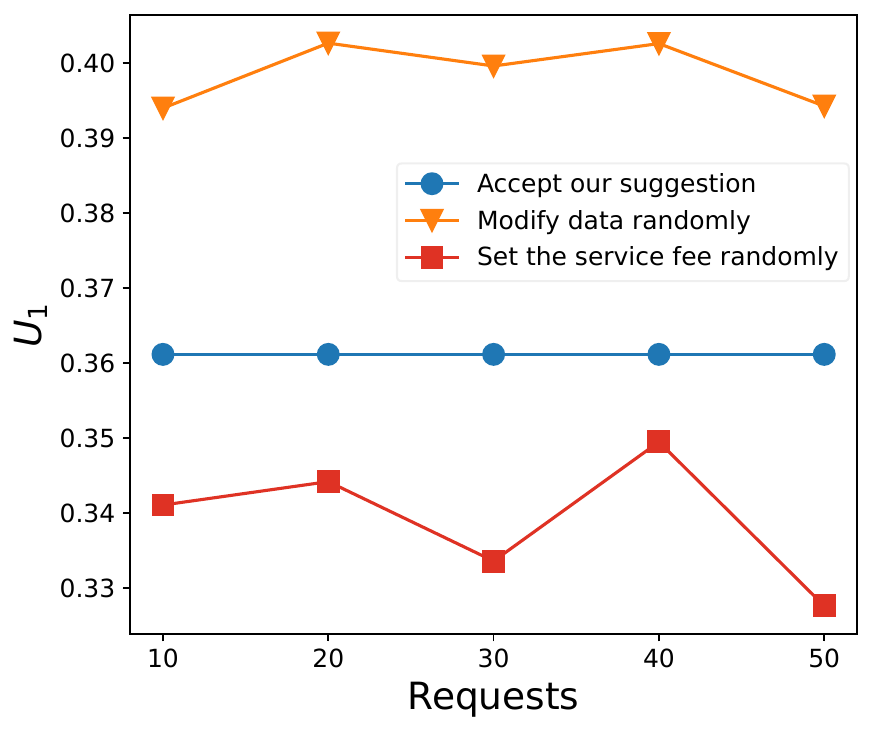}}\quad
\subfloat[Benefits for task executors]{\label{fig:u2}\includegraphics[width=0.45\linewidth]{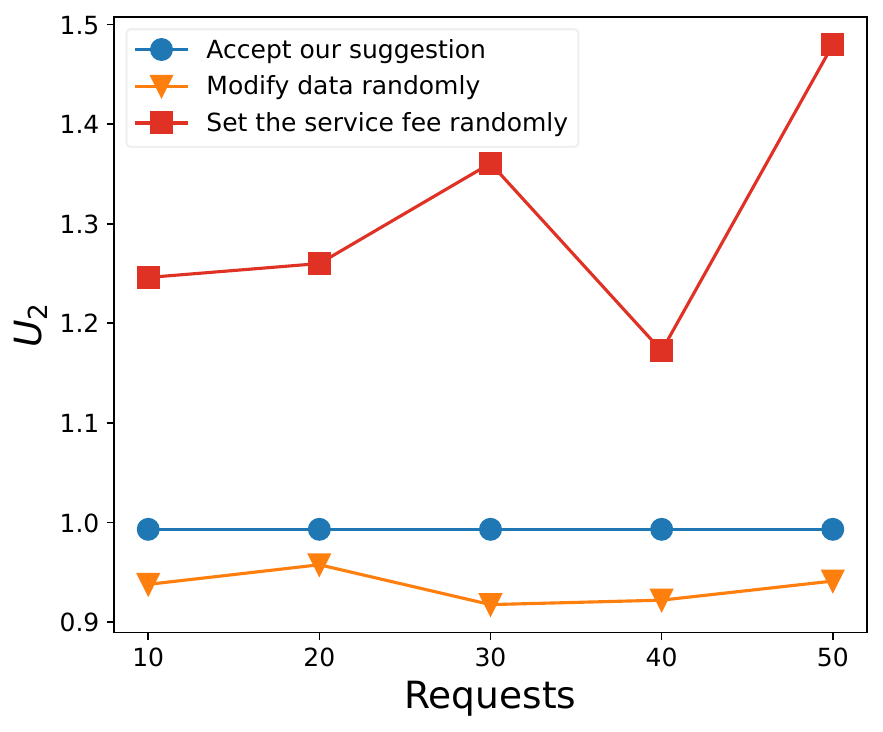}}\\
\caption{Benefits of task publishers and executors.}
\label{fig:benefit1}
\end{figure}

\begin{figure*}[h]
\centering
\subfloat[Relationship between $P$ and $U_2$]{\label{fig:p_u2}\includegraphics[width=0.3\linewidth]{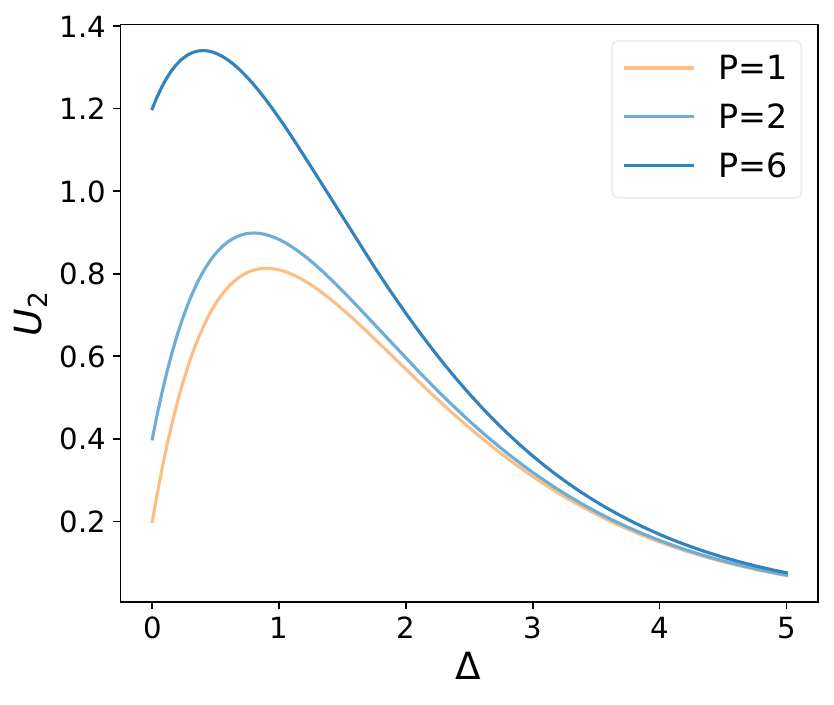}}\quad
\subfloat[Relationship between $\alpha$ and $P$]{\label{fig:k_p}\includegraphics[width=0.3\linewidth]{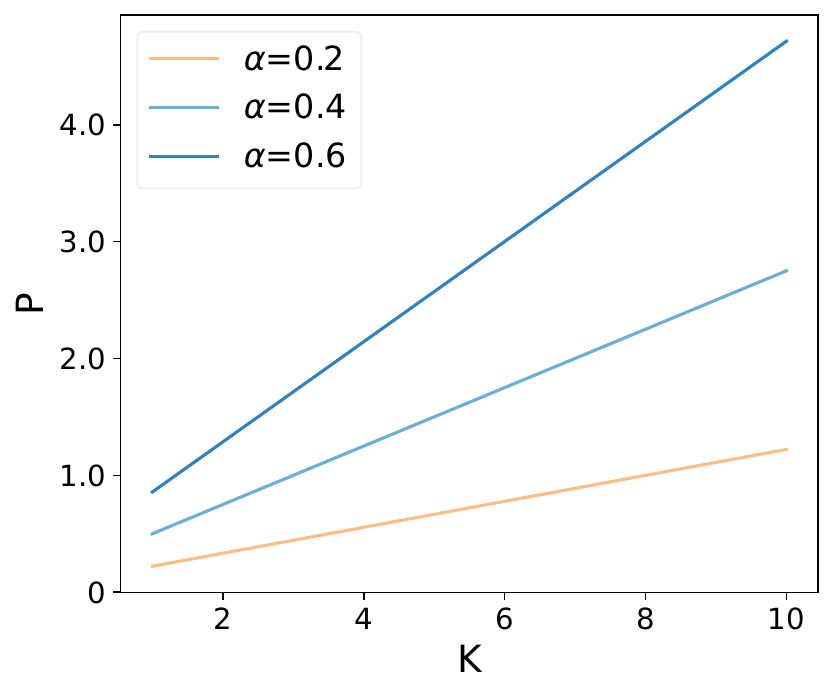}}\quad
\subfloat[Relationship between $\alpha$ and $\Delta$]{\label{fig:k_delta}\includegraphics[width=0.3\linewidth]{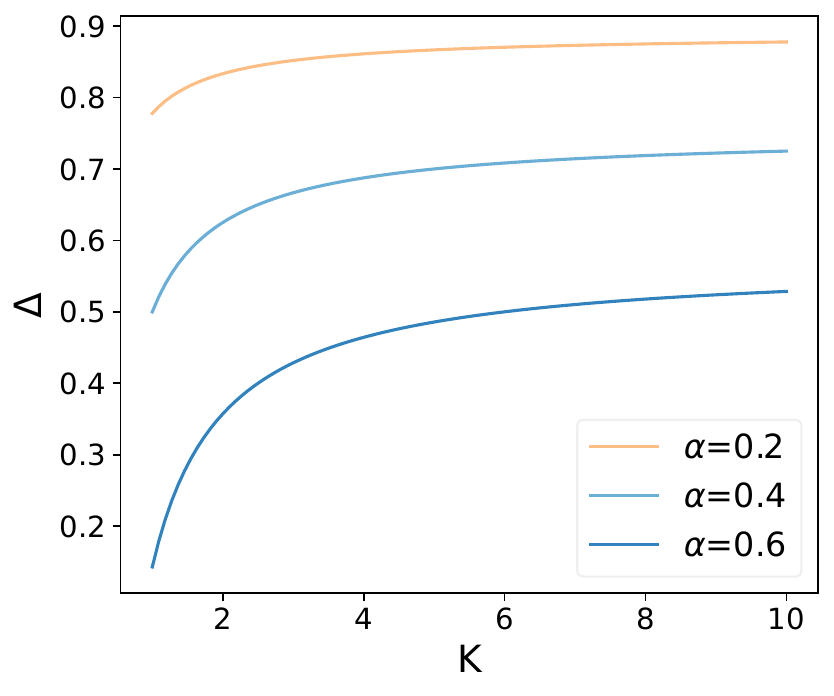}}\\
\caption{The impact of various parameters in incentive mechanisms on user behavior.}
\label{fig:benefit2}
\end{figure*}

Three experiments in Fig. \ref{fig:select_node} show the advantages of the proposed scheme over Baseline in terms of quality of service. The proposed node selection scheme can select higher-quality nodes for tasks while ensuring the anonymous security of nodes, and effectively improving the service quality of the system.

\subsection{Incentive Mechanism}
Fig. \ref{fig:benefit1} compares the profits obtained by the task publisher and the malicious task executor when they reject our recommendations in 50 tasks. When they reject our suggestions, i.e., when the task publisher sets service fees randomly and the task executor modifies data randomly, the profits obtained are always lower than when they accept our recommendations. Taking the task publisher as an example, when they fail to adopt the optimal strategy of the game to set the service fee but set the service fee randomly, the profit $U_1$ obtained decreases. Similarly, this holds for the executor. Therefore, for rational participants, the behavior that maximizes their benefits is also to follow the equilibrium solution given by us.

Fig. \ref{fig:p_u2} illustrates the impact of the service fee $P$ set by the task publisher on the earnings and behavior of the executor. When the task publisher sets a higher $P$, the executor obtains higher revenues $U_2$ and tends to engage in honest behavior, making fewer modifications $\Delta$ of data. However, it is not necessarily advantageous for the publisher to pay a higher fee of $P$. The publisher expects to spend the minimum fee while obtaining data that meets its requirements. As expected, the task publisher can influence the behavior of the executor by setting $P$.

Fig. \ref{fig:k_p} illustrates the relationship between the task publisher's expectation of data quality, represented by $\alpha$, and the optimal service fee $P$. Consistent with expectations, as $\alpha$ increases, the publisher tends to set a higher fee to obtain higher-quality data. Additionally, $P$ is positively correlated with $K$, indicating that as the number of goods to be traded/evaluated increases, malicious executors become more profitable, requiring higher fees to ensure data quality. Fig. \ref{fig:k_p} demonstrates that the proposed incentive mechanism can adjust the recommended optimal service fee $P$ according to the data requirements of the task publisher.

Fig. \ref{fig:k_delta} shows the impact of the task publisher's expected data quality requirement $\alpha$ on the malicious executor's modification of the data range $\Delta$. We find that as $\alpha$ increases, $\Delta$ decreases, and the executor tends to modify the data less and slowly increases as $K$ increases. It confirms the effectiveness of the proposed incentive mechanism from the side and can return results that meet the expected requirements according to the needs of the task publisher.

Fig. \ref{fig:benefit1} verifies that the proposal given by the proposed scheme is the optimal solution for each participant to obtain benefits. Moreover, we analyze the effectiveness of the proposed incentive mechanism in detail in Fig. \ref{fig:benefit2}, including the impact of the $P$ set by the task publisher on the task executor, and the impact of the task publisher's expected $\alpha$ on its pricing and final data offset.

\begin{figure}[h!]
\centering
\subfloat[Robustness]{\label{fig:safe}\includegraphics[width=0.45\linewidth]{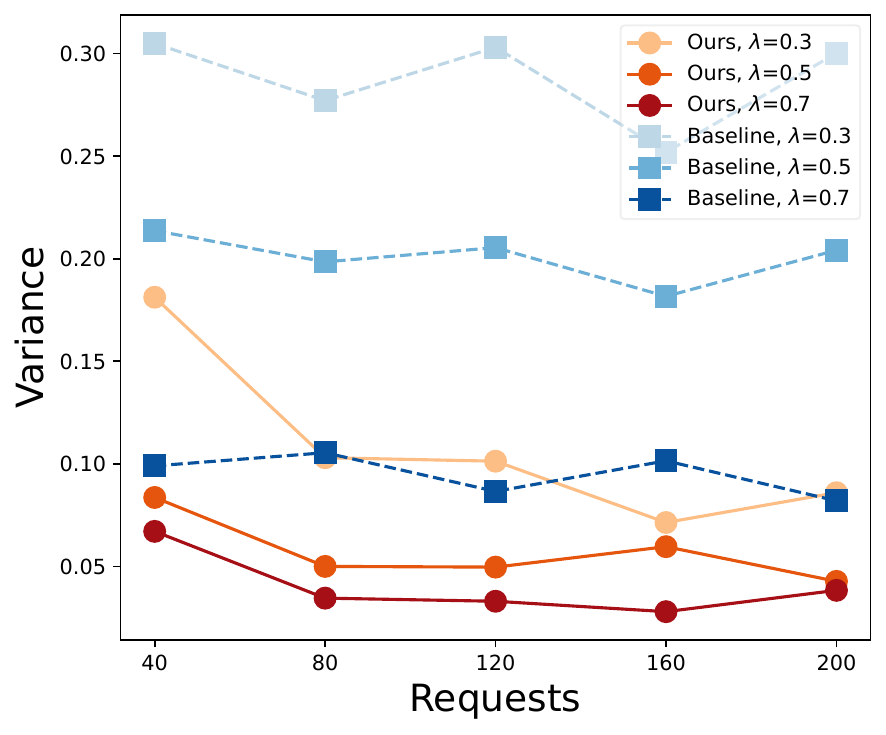}}\quad
\subfloat[Scalability]{\label{fig:scalability}\includegraphics[width=0.45\linewidth]{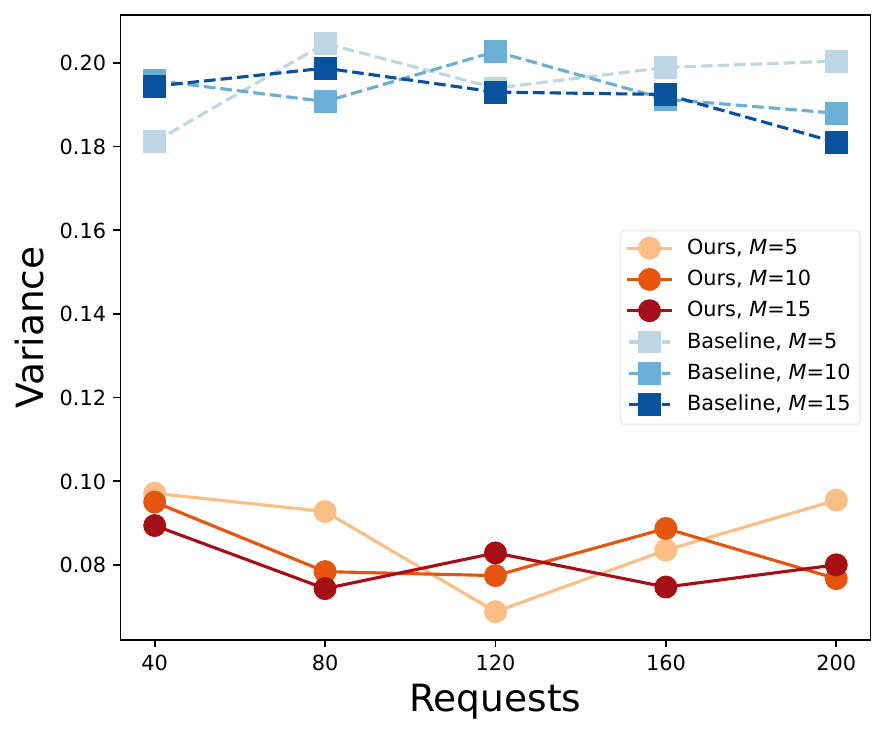}}\\
\caption{The robustness and scalability of the proposed scheme.}
\label{fig:robustness}
\end{figure}

\subsection{Robustness and Scalability}

Fig. \ref{fig:safe} verifies the robustness of the proposed scheme in the oracle network with different proportions of malicious nodes $\lambda$. Although, with the increase in the proportion of malicious nodes, the consistency of data is declining. However, compared with Baseline, our scheme can maintain good performance in various environments. It shows that the proposed scheme has good robustness.

Fig. \ref{fig:scalability} verifies the scalability of the proposed scheme. It can be found that the increase in the number of nodes $M$ selected in a task will not have much effect on the consistency of the data obtained. Moreover, compared with Baseline, our scheme can maintain higher data consistency. It shows that the proposed scheme has good scalability.

\section{Security Analysis}
\label{security_analysis}

We briefly analyze the security of the proposed scheme in the face of the Sybil attack, Targeted attack, and Collusion attack.

\paragraph{Sybil Attack}
First of all, similar to conventional schemes such as ChainLink \cite{chainlink}, this scheme can effectively limit the disorderly expansion of nodes by setting a token pledge mechanism in the registration contract. Secondly, the proposed incentive mechanism improves the base number of network nodes, the reputation mechanism added in the node selection strategy reduces the selected weight of malicious nodes, and the overall security is improved.

\paragraph{Targeted Attack}
Targeted attacks are not a specific modality of assault, they denote a sustained assault by the aggressor against a specific target, such as the task node selected for the current task \cite{sun2020voting}. This approach inherits the unpredictability yet verifiability from the VRF, ensuring anonymity security in the node selection process under the assumption of the computational hardness of the discrete logarithm problem. Furthermore, the scale effect of the oracle network also reduces the probability of attackers targeting task executors.

\paragraph{Collusion Attack}
Collusion attack refers to multiple attackers conspiring and collaborating to carry out a particular attack \cite{li2020rational}. On one hand, the anonymity in the node selection process can increase the difficulty of collusion among nodes to some extent. On the other hand, enlarging the scale of the network can effectively reduce the probability of their attack success.

In summary, we can observe that the scale of the oracle network is one of the key factors influencing its security. Incentivizing more users to join the oracle network can effectively enhance its resilience against external attacks. However, merely meeting the revenue of oracle nodes is not sustainable, it is also necessary to ensure that task publishers are willing to participate to promote the system's sustainable development and enhance its risk resistance capabilities.

\section{Conclusion}
\label{conclusion}
This paper proposes an oracle scheme to provide reliable price data for applications such as payment and settlement, and asset valuation on the blockchain. To ensure the security of nodes and improve the consistency of returned data, we design an anonymous node selection scheme based on the reputation mechanism and VRF. Then, to meet the revenue of all parties, we propose an incentive mechanism based on the Stackelberg game to provide optimal service fee-setting suggestions for task publishers. Experiments and security analysis show that the scheme can improve data consistency while satisfying anonymous security and mutual revenue. In the future, we will continue to refine model details, expand data types, optimize data filtering mechanisms, etc.

\appendices
\section{Solve the optimal decision of follower}
\label{proof1}

The goal of the follower is to maximize their revenue $U_2$, that is, to solve the maximum value of $U_2$:

\begin{eqnarray}
\frac{d U_2}{d \Delta} = \frac{(-\Delta K+K-P)\times e^{-\Delta}}{n}
\end{eqnarray}

Let $\frac{d U_2}{d \Delta} =0$:
\begin{eqnarray}
 \Delta = \frac{K-P}{K}
\end{eqnarray}

At the same time:
\begin{eqnarray}
\frac{d^2 U_2}{d \Delta^2} =  \frac{-(K -\Delta K + K-P)\times e^{-\Delta}}{n}
\end{eqnarray}

If follower wants to obtain the maximum value of $U_2$, then $\frac{d^2 U_2}{d \Delta^2} < 0$, so:
\begin{eqnarray}
(2-\Delta) K - P > 0 
\end{eqnarray}

$\Delta = \frac{K-P}{K}$ is brought into $\frac{d^2 U_2}{d \Delta^2}$, which is always less than 0, so $\Delta^{*} = \frac{K-P}{K}$ is the optimal decision of the follower when the $P$ set by the leader is known.

\section{Solve the optimal decision of leader}
\label{proof2}

Under the condition that the optimal strategy $\Delta^{*} = \frac{K-P}{K}$ of the follower is known, the leader takes $\Delta^{*}$ into his own revenue function $U_1$ to obtain the optimal solution $P^{*}$:

\begin{eqnarray}
\frac{d U_1}{d P} = \frac{(K(\alpha-1)+P(\alpha-1)+\alpha)e^{\frac{P-K}{K}}}{K}
\end{eqnarray}

Let $\frac{d U_1}{d P} = 0$:
\begin{eqnarray}
P = \frac{K(\alpha-1) + \alpha}{(1-\alpha)}
\end{eqnarray}

Leader wants to get the maximum value of $U_1$, then $\frac{d^2 U_1}{d P^2} < 0$, so:
\begin{eqnarray}
2K(\alpha-1)+P(\alpha-1)+\alpha < 0
\end{eqnarray}

Bring $P = \frac{K(\alpha-1) + \alpha}{(1-\alpha)}$ into it, we get $K(\alpha-1) < 0$.

Because $K>0$ and $\alpha < \frac{2K}{1+2K} < 1$, so $\frac{d^2 U_1}{d P^2} < 0$ always holds. And $0 \leq \Delta $ and $0 \leq P$ is known, so $0 \leq P \leq K$, so:
\begin{eqnarray}
\frac{K}{1+K} \leq \alpha \leq \frac{2K}{1+2K}
\end{eqnarray}

Therefore, when $\frac{K}{1+K} \leq \alpha \leq \frac{2K}{1+2K}$, Leader chooses to set the optimal service price $P^{*} = \frac{K(\alpha-1) + \alpha}{(1-\alpha)}$ to obtain the maximum value of its interest $U_1$.

\bibliographystyle{IEEEtran}
\bibliography{IEEEabrv,myref}

\end{document}